\documentclass{elsart}

 \usepackage{graphicx}

\usepackage{amssymb}
\usepackage{amsmath}
\usepackage{amsfonts}%

\begin{document}

\begin{frontmatter}



\title{
Current-induced spin polarization and the spin Hall effect:
a quasiclassical approach}


\author[rome]{R. Raimondi},
\author[augsburg]{C. Gorini},
\author[augsburg]{M. Dzierzawa},
\author[augsburg]{P. Schwab}

\address[rome]{Dipartimento di Fisica "E. Amaldi", Universit\`a di Roma Tre, Via della Vasca Navale 84, 00146 Roma, Italy}
\address[augsburg]{Institut f\"ur Physik, Universit\"at Augsburg, 86135 Augsburg, Germany}
\begin{abstract}
The quasiclassical Green function formalism is used to describe charge and spin dynamics 
in the presence of spin-orbit coupling. We review the results obtained for the spin Hall
effect on restricted geometries. The role of boundaries is discussed
in the framework of spin diffusion equations.

\end{abstract}

\end{frontmatter}



\section{Introduction}
\label{}
A transverse (say along the $y$-axis) $z$-polarized spin current
flowing
in response to an applied (say along the $x$-axis) electrical field
\begin{equation}
\label{eq1 }
J^z_{s,y} = \sigma_{sH} E_x,
\end{equation}  
is referred to as the spin Hall effect, and
$\sigma_{sH}$ is called spin Hall conductivity\cite{murakami2003,sinova2004}.  
This effect allows for the generation and control of spin currents by
purely electrical means, which is a great advantage when operating electronic devices. Physically, the possibility of a non-vanishing 
$\sigma_{sH}$ is due to the presence of spin-orbit coupling, which in semiconductors
may be orders of magnitude larger than in vacuum. Clearly, two key issues are (i) how sensitive is the spin
Hall conductivity to various solid state effects like disorder scattering, 
electron-electron interaction etc.{} and (ii) how the spin current can actually be detected experimentally.  About the first
issue, there is now a consensus that the effect of disorder scattering depends on the form of the spin-orbit coupling. 
In the case of the two-dimensional electron gas 
with a Rashba type of spin-orbit coupling, arbitrarily weak disorder leads to the vanishing of  the spin Hall conductivity in the bulk 
limit\cite{mishchenko2004,inoue2004,raimondi2005,dimitrova2005,chalaev2005, murakami2004,khaetskii2006}. Concerning the second issue, the first
experimental observations\cite{kato2004,wunderlich2005} of the spin Hall effect have been achieved by measuring, optically, 
the spin polarization accumulated at the lateral edges of an
electrically biased wire. Hence, the understanding of the spin Hall effect involves the description of boundaries. In order 
to address the two issues mentioned above, we develop in the following
a quasiclassical Green function approach for the description of charge and spin degrees of freedom in the presence of
spin-orbit coupling.


\section{The Eilenberger equation}
In this section we sketch how to derive the Eilenberger equation for the quasiclassical Green function 
in the presence of spin-orbit interaction \cite{raimondi2006}.
We consider the following  Hamiltonian 
\begin{equation}
\label{eq2}
H = \frac{p^2}{2m}+\mathbf{ b}\cdot {\boldsymbol \sigma},
\end{equation}
where $\textbf{b}(\mathbf{p})$ is an effective internal magnetic field due to the spin-orbit coupling.
For  instance, in the Rashba case we have $\mathbf{ b}=\alpha \mathbf{ p}\times\mathbf{ {\hat e}}_z$.
The Green function $\check{G}_{t_1t_2}(\textbf{x}_1 ,\textbf{x}_2)$ 
is a matrix both in Keldysh and spin space and obeys the Dyson equation $(\hbar = 1)$
\begin{equation}
\label{eq3}
\left(\textrm{i}\partial_{t_1}+\frac{1}{2m}\partial^2_{\mathbf{x}_1}+\mu -
\mathbf{ b}(-\textrm{i}\partial_{\mathbf{x}_1})\cdot {\boldsymbol \sigma}  \right)\check{G}_{t_1 t_2}(\mathbf{ x}_1 ,\mathbf{ x}_2)=\delta (t_1-t_2)\delta (\mathbf{ x}_1 -\mathbf{ x}_2).
\end{equation}
A quasiclassical description is possible when the Green function depends
on the center-of-mass coordinate, $\mathbf{ x}=(\mathbf{ x}_1+\mathbf{ x}_2)/2$, on a much larger scale 
than on the relative coordinate, ${\mathbf r}=\mathbf{ x}_1-\mathbf{ x}_2$. In this situation, 
by going to the Wigner representation
\begin{equation}
\label{ eq4}
\check{G}(\mathbf{ x}_1, \mathbf{ x}_2)=\sum_{\mathbf{ p}}e^{\textrm{i}\mathbf{ p}\cdot {\mathbf r}}
\check{G}(\mathbf{ p},\mathbf{ x)}
\end{equation}
and subtracting from Eq.{}(\ref{eq3}) its complex conjugate, one obtains a homogeneous equation for $\check{G}(\mathbf{p},\mathbf{x})$
\begin{equation}
\label{eq5}
\textrm{i}\partial_t \check{G}+\frac{\textrm{i}}{2}
\big\{\frac{\mathbf{ p}}{m}+\frac{1}{2}\partial_{\mathbf{ p}}(\mathbf{ b}\cdot {\boldsymbol \sigma}),
\partial_\mathbf{ x}\check{G}\big\}-\left[\mathbf{ b}\cdot {\boldsymbol \sigma}, \check{G}\right]=\left[\Sigma, \check{G}\right],
\end{equation}
where only the center-of-mass  time, $t=(t_1+t_2)/2$, appears. 
On the right-hand-side of Eq.{}(\ref{eq5}) we have also introduced the self-energy
\begin{equation}
\label{eq5b}
\Sigma =\frac{1}{2 \pi N_0 \tau}\sum_{\mathbf{p}}\check{G}(\mathbf{p},\mathbf{x}),
\end{equation}
which takes into account disorder scattering at the level of the self-consistent 
Born approximation. 
In the spirit of the quasiclassical approximation, we make the following ansatz for $\check{G}$
\begin{equation}
\check G = 
\left( \begin{array}{cc} 
G^R & G^K \\
0   & G^A 
\end{array}
\right)
= \frac{1}{2} \left\{ 
\left( \begin{array}{cc} 
G_0^R & 0 \\
0   & -G_0^A 
\end{array}
\right),
\left( \begin{array}{cc} 
\tilde g^R & \tilde g^K \\
0   & \tilde g^A 
\end{array}
\right)
\right\}
,\end{equation}
where we assume that $\tilde g$ does not depend on the modulus of $\mathbf p$ but at most
on its direction. In this way 
we have separated the fast variation of the free Green function in the relative coordinate $\mathbf r$
from the slow variation of $\tilde g$ in the center-of-mass coordinate $\mathbf x$.
The quasiclassical Green function is defined as
\begin{equation}
\label{eq8}
\check{g}(\mathbf{\hat{p}},\mathbf{x})\equiv \frac{i}{\pi}
\int_{-\infty}^{\infty}\mathrm{d}\xi\ \check G(\mathbf{p},\mathbf{x})
,\end{equation}
where $\xi =p^2/2m -\mu$ and $\mathbf{\hat{p}} = \mathbf{p}/p$.
Using the ansatz above, the $\xi$-integration
can be done explicitly.
By assuming that $\mathbf{b} = b(p)\mathbf{\hat{b}}(\mathbf{\hat{p}})$  and
\begin{equation}
G_0^{R}=
\sum_{\nu = \pm }\frac{\textbf{P}_{\nu}}{\epsilon-\xi-\nu b (\xi ) + \mathrm{i}0^+}, \ \
\textbf{P}_\pm =\frac{1}{2}\left(1 \pm \hat{\textbf{b}}\cdot {\boldsymbol \sigma} \right)
\end{equation}
we find
\begin{equation}
\label{eq7}
\frac{i}{\pi}\int_{-\infty}^{\infty}\mathrm{d}\xi \ G_0^R =\frac{N_+}{N_0}\textbf{P}_++\frac{N_-}{N_0}\textbf{P}_-
\equiv g_0^R
,\end{equation}
where $N_+$ and $N_-$ are the densities of state of the spin-split
bands.
In the absence of spin-orbit coupling, $N_{\pm}=N_0$, 
and the function $\tilde{g}$ coincides with the  $\xi$-integrated Green function $\check{g}$. In the present case,  
however,
\begin{equation} \label{8}
 \check g   = \frac{1}{2} \{ g_0^R ,\tilde{g}\}.
\end{equation}
By inverting Eq.{}(\ref{8}) one has
\begin{equation}
\label{eq10}
\tilde{g}=\frac{1}{2}\{(g_0^R)^{-1},g\}+
          \frac{1}{4}\left[(g_0^R)^{-1},\left[ g_0^R,g\right]\right]
\approx \frac{1}{2}\left\{
  \left( 1  - \frac{N_+-N_-}{2N_0}\hat{\textbf{b}} \cdot   {\boldsymbol \sigma} \right),
  \check{g}\right\},
\end{equation}
where the last approximation, to be used in the following, holds for
$b \ll \epsilon_F$. 
The  $\xi$-integration of $\check{G}$ multiplied by a momentum dependent function leads to
\begin{equation}
\label{eq9}
\frac{i}{\pi}\int_{-\infty}^{\infty}\mathrm{d}\xi \ m (\textbf{p})
\check G =
\frac{1}{2} \{\textbf{M} g_0^R ,\tilde{g}\}.
\end{equation}
where $\textbf{M}= m(\textbf{p}_+)\textbf{P}_++
m(\textbf{p}_-)\textbf{P}_-$, and ${\bf p}_\pm$ are the Fermi momenta
in the two bands.
To first order in $b /\epsilon_F$ one obtains
\begin{equation}
\label{eq11}
 \frac{1}{2} \{\mathbf{ M }g_0^R,\tilde{g}\}
\approx \frac{1}{2}\{\mathbf{ M},\check{g}\}
= m(\textbf{p}_+)\check{g}_++m(\textbf{p}_-)\check{g}_-
,\end{equation}
where $\check g_\pm = \frac{1}{2} \{P_\pm , \check g \}$. 
Hence the standard procedure to obtain the Eilenberger equation  
for $\check{g}$  via the $\xi$-integration of Eq.{}(\ref{eq5}) yields
\begin{equation}
\label{eq12}
\sum_{\nu = \pm }\big(
  \partial_{t} \check g_\nu  + 
  \frac{1}{2} \left\{
    \frac{\bf p_\nu}{m}  
   +\frac{\partial}{\partial \bf p}({\bf b}_\nu \cdot {\boldsymbol  \sigma}),
    \frac{\partial}{\partial {\bf x} }\check g_\nu \right\} 
+  {\rm i } [{\bf b}_\nu \cdot {\boldsymbol \sigma}, \check
g_\nu ]\big)
= - {\rm i} \left[ \check \Sigma , \check g \right].
\end{equation}
Eq.{}(\ref{eq12}) holds even for internal fields $\mathbf b$ for which the
factorization $\mathbf{b} = b(p)\mathbf{\hat{b}}(\mathbf{\hat{p}})$ is not possible, 
as long as one can assume $|\mathbf b| \ll \epsilon_F$. For more details see \cite{raimondi2006}.

From the above Eilenberger equation the expression for the charge and spin currents
are readily obtained, since the Fermi surface average $\langle \ldots \rangle $ of 
Eq.(\ref{eq12}) has the form of 
a continuity equation
\begin{eqnarray}\label{eq13}
\partial_t \langle \check g_c \rangle + \partial_{\bf x} \cdot \check {\bf J}_{c} &=&0 \\
\label{eq14}
\partial_t \langle \check g_x \rangle + 
\partial_{\bf x} \cdot\check {\bf J}_{s}^x&=&2
\sum_{\nu = \pm }  \langle {\bf b}_\nu \times {\bf \check g_\nu} \rangle_x \\
\label{eq15}
\partial_t \langle \check g_y \rangle + 
\partial_{\bf x } \cdot \check {\bf J}_{s}^y&=&2
\sum_{\nu = \pm }  \langle {\bf b}_\nu \times {\bf \check g_\nu} \rangle_y \\
\label{eq16}
\partial_t \langle \check g_z \rangle + 
\partial_{\bf x}\cdot \check {\bf J}_{s}^z  &=&2
\sum_{\nu = \pm }  \langle {\bf b}_\nu \times {\bf \check g_\nu} \rangle_z ,
\end{eqnarray}
where we expanded the Green function in charge and spin components,
$\check g=\check g_c + \check{ \bf g}\cdot {\boldsymbol \sigma} $.
For instance, the spin current with spin polarization along the $\mathbf{ e}_z$ axis reads
\begin{equation}
\label{eq17}
{\bf j}_s^z({\bf x}, t) = - \frac{N_0}{4} \int {\rm d }\epsilon \,  [
{\bf {\check J}}_s^z(\epsilon; {\bf x}, t) ]^K, 
\end{equation}
with  
${\check  {\bf  J}}_s^z(\epsilon; {\bf x}, t)  = \langle {\bf v}_F {\check
g}_z  \rangle$.


\section{Spin Hall effect}
Before considering explicitly the role of the boundaries, 
it is useful to see how the spin Hall effect in the bulk may be analyzed in the present formalism. 
For a stationary and space independent case the equation for the Keldysh component
can be written in the following way ($a= 2 \alpha p_F \tau$)
\begin{equation}
\label{18}
\left(
\begin{array}{cccc}
     1       &   0   &   0   & 0 \\
     0       &   1   &   0   & a \hat p_x \\
     0       &   0   &   1   & a\hat p_y \\
     0       & -a  \hat p_x & -a  \hat p_y & 1
   \end{array}
 \right) 
 \left(
\begin{array}{c}
 g_c^K \\
 g_x^K \\
 g_y^K \\
 g_z^K
\end{array}
\right)
=\left(
\begin{array}{cccc}
   1    & -{\alpha }\hat{ p }_y /v_F & {\alpha}\hat{p}_x /v_F & 0 \\
     -{\alpha} \hat{p}_y/v_F &   1    &   0 &  0 \\
        {\alpha} \hat{p}_x/v_F &   0    &   1 &  0 \\
           0   &   0    &  0  & 1
           \end{array}
           \right)
           \left(
           \begin{array}{c}
           \langle g_c^K \rangle \\
           \langle g_x^K \rangle \\
           \langle g_y^K \rangle\\
           \langle g_z^K \rangle
               \end{array}
               \right)
           + S_E
.\end{equation}
In the above we have introduced the mean free path $l=v_F \tau$. 
The presence of the electric field, along the $\mathbf{ {\hat e}}_x$ axis,  is accounted for by the source term $S_E$
\begin{equation}
\label{19}
S_E =  - 4 |e|E l f'(\epsilon)
\left[
\left(
\begin{array}{c}
\hat{ p}_x \\
0 \\
0 \\
0 \\
\end{array}
\right)
+  \frac{\alpha}{v_F}
\left(
\begin{array}{c}
0 \\
-2 \hat{ p}_x \hat{ p}_y \\
\hat{ p}_x^2 - \hat{ p}_y^2 \\
0 \\
\end{array}
\right)
\right],
\end{equation}
which has been introduced in the Eilenberger equation by exploiting the gauge invariance in the derivative terms.  In the above $f(\epsilon)$ is the Fermi function.  
Notice that spin-charge coupling effects arise at the order of $\alpha /v_F$.
From the angular average of Eq.(\ref{18}) one observes immediately
that $\langle {\hat p}_x g_z^K \rangle = \langle {\hat p}_y g_z^K  \rangle =0$, i.{}e.{} no
spin current with polarization along  ${\bf  {\hat e}}_z$ flows in the
system.
It is also instructive to first express $g^K_z$ in terms of the angle-averaged 
quantity $\langle g^K \rangle$ and then multiply by $p_y$ and take the
angular average. 
The spin current reads then
\begin{equation}
\label{eq20}
j_y^z = 
\frac{v_F \alpha p_F \tau}{1 + (2\alpha p_F \tau)^2}
\left( 
\alpha | e | \tau N_0 E + s_y \right)
.\end{equation}
When comparing to a calculation of the spin-Hall conductivity within the standard
Kubo-formula approach, one realizes that 
the first term in square brackets corresponds to the simple bubble 
diagram, while the second term 
accounts for vertex corrections due to the spin-dependent part of the velocity operator. 
This second term 
is related to a voltage induced spin polarization in ${\bf {\hat e}}_y$
direction that was first obtained by
Edelstein \cite{edelstein}, $ s_y = - |e| \alpha \tau N_0 E$.

In summary, we find that under the very special conditions we
considered here the spin-Hall current is zero. 
We started from the Rashba Hamiltonian, that
has a linear-in-momentum spin orbit coupling, we assumed that elastic
scattering is spin-conserving, and finally we considered the
stationary solution in the bulk of the two-dimensional electron
system so that the spin density neither depends on space nor time.
Relaxing one or more of these conditions finite spin-Hall currents are
expected.


\section{Boundary conditions and diffusion equation}

The derivation of the boundary conditions for the quasiclassical 
Green function is a delicate task, due to the fact that the boundary potential 
typically varies on a microscopic length scale which is shorter than the 
quasiclassical \textsl{resolution}. Therefore one must include the boundary effect in the very process 
of the derivation of the Eilenberger equation. 
This is especially important for interfaces 
between two different materials, where transport occurs through a tunnel junction or a point contact. 
Often a boundary with vacuum is simpler to describe since one couples
only one incoming with one outgoing channel.
In the presence of spin-orbit coupling, 
the non-conservation of spin and in particular 
the beam splitting, i.e., one incoming channel with direction ${\bf p}_{in}$ can be
scattered into two outgoing channels ${\bf p}_{out}$,
makes the problem more difficult and there is not yet a 
complete derivation of boundary conditions for the quasiclassical Green function.

In the following we limit to two special types of boundary conditions
and discuss their role as far as spin relaxation and the spin Hall effect are 
concerned\cite{schwab2006}.
The first type of boundary conditions requires spin conservation,
i.e.{} the spin currents normal to  
the interface are zero. We refer to them in the following as {\it hard-wall}
boundary conditions, (see also \cite{galitski2006,bleibaum2006}).
For smooth confining potential it has been pointed out, that specular scattering 
involves some spin rotation in such a way to keep the scattered particle in the eigenstate
of the incoming one \cite{govorov2004}. We refer to this situation as {\it soft-wall} boundary conditions.

The matching condition for the quasiclassical Green function can be
obtained following the approach of Ref. \cite{millis1988}
and reads
\begin{equation}
 \label{22}
 \check{g}(\mathbf{ \hat{p}}_{out} )= S \check{g}(\mathbf{ \hat{p}}_{in})S^{\dag}
,\end{equation}
where the matrix $S$ describes the scattering at the boundary.

To illustrate the role of boundaries, we consider the 
diffusive limit where a detailed analytical treatment is possible and which is also relevant
for actual experiments.  
\begin{figure}
\centerline{\includegraphics[width=0.5\textwidth]{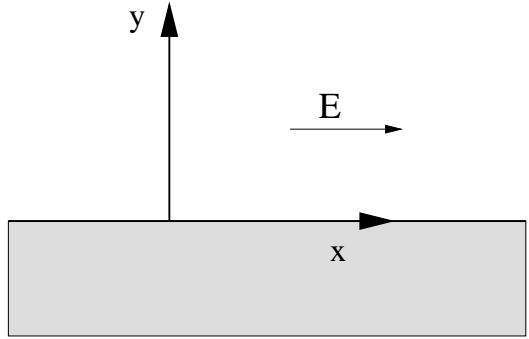} }
\caption{\label{fig1} 
Orientation of the boundary and the electric field.}
\end{figure}
We assume that our two-dimensional electron gas is limited to the 
half-space $y>0$ and that there is a uniform electric field along the
$\mathbf{ \hat{e}}_x$ direction, see Fig. \ref{fig1}. 
Since the system is translational invariant along the $\mathbf{ \hat{e}}_x$ direction,  
we consider only space dependence with respect to the $\mathbf{ \hat{e}}_y$
direction. The diffusion equation reads
\begin{eqnarray} \label{23}
\left( \partial_t -  D \partial_{ y}^2  \right) \rho & = & 2B\partial_y s_x \\
\label{24}
\left( \partial_t -  D \partial_{ y}^2 \right) s_x   & = &   -
\frac{1}{\tau_s}   s_x
+2B \partial_y \rho  \\
\label{25}
\left( \partial_t -  D \partial_{y}^2 \right) s_y    &= &
-\frac{1}{\tau_s}  ( s_y -s_0) 
+2C \partial_y s_z  \\
\label{26}
\left( \partial_t -  D \partial_{y}^2 \right) s_z &= &  -
\frac{2}{\tau_s}   s_z
  - 2 C \partial_y s_y  
,\end{eqnarray}
with $D=\frac{1}{2}v_F^2 \tau$, $\tau_s = \tau/[2(\alpha p_F \tau
)^2]$, $B=4 \alpha^3 p_F^2 \tau^2$,  $C= v_F \alpha p_F \tau$, and
$s_0= - |e|\alpha \tau N_0 E$ is the bulk spin polarization in the presence of an electric field,
 mentioned at the end of the previous section. 
A general solution of the diffusion equations (\ref{23}-\ref{26}) can be found in the form
\begin{equation}
\label{27}
 s(y,t)=e^{-\gamma t} e^{\textrm{i}q y}
\left(\begin{array}{c}\rho \\ s_x \\s_y \\s_z \end{array}\right),
\end{equation}
where a static solution requires $\gamma =0$. 
 In the absence of electric field (i.e., $s_0=0$), as in optical spin excitation experiments, the solution with longest lifetime has a finite wave vector in contrast to
standard diffusion. The presence of boundaries does affect this result.
With hard-wall boundary conditions
\begin{eqnarray} 
   \label{28}
   -D \partial_y  s_x   =      {\bf n} \cdot
   {\bf j}_x  & = & 0 , \\
   \label{29}
   -D \partial_y  s_y    - C s_z 
    =    {\bf n} \cdot {\bf j}_y      & = &0 ,\\
   \label{30}
   - D  \partial_y  s_z   + C  (s_y -s_0)    =  {\bf n} \cdot
   {\bf j}_z & = &0
,\end{eqnarray}
where  ${\bf n}$ is in the $y$-direction, the mode with longest lifetime is localized at the boundary.
With  soft-wall boundary conditions, one has 
\begin{equation} \label{31}
s_x=0 \, \, \text{and}  \, \, s_y=s_0,
\end{equation}
while the $z$-component of the
spin is still conserved and therefore Eq.\ (\ref{30}) remains valid. 
Due to the decoupling, at the boundary, of the three spin components, the mode
with longest lifetime is no longer confined to the boundary and 
has $s_{x}\propto 	\sin (q y)$, $s_y-s_0 \propto \sin (q y)$, $s_z \propto \cos (q y)$ with $q\sim L_s^{-1}$ (as in the bulk),
where $L_s =\sqrt{D\tau_s}$ is the spin diffusion length.
Although the two types of boundary conditions have a different effect in a time-dependent optical spin-relaxation experiment, both imply that, in the absence of an electrical field, 
 the only static solution  of the diffusion equation is with vanishing polarization  (See Ref.\cite{schwab2006} for further details).

 The presence of an electric field does not change this result provided the polarization in the $\mathbf{ \hat{e}}_y$ direction is replaced by the difference $s_y -s_0$, which enters both the diffusion equations  (\ref{23}-\ref{26}) and
the  boundary conditions (\ref{28}-\ref{31}). Hence, in the presence of an electric field, the only static solution has $s_y=s_0$, $s_x=0$ and $s_z=0$. 
Furthermore, the second condition in (\ref{31}) means that, with soft-wall boundary conditions and  at the level of diffusive accuracy, the time-dependent approach to 
the static solution  becomes infinitely fast  at the boundary,
in contrast to what happens in the bulk where relaxation occurs in a finite time.


\section{Numerical results}

In this section we give a brief overview of numerical results for both spin Hall effect and spin accumulation in finite systems. In the following we consider 
the Rashba Hamiltonian on a rectangular strip of length $L_x$ and width $L_y$ which
is connected to reservoirs at $x=0$ and $x=L_x$. We impose soft-wall boundary conditions at 
$y = 0$ and $y = L_y$. At $t = 0$ the electrical field $E$ is applied in $x-$direction.  
In order to solve the time-dependent Eilenberger equation (\ref{eq12}) numerically we introduce
the discretization $\Delta x$ and $\Delta t$ for space and time, respectively.
Our algorithm is exact to order $(\Delta t)^2$ and $(\Delta x)^4$ and yields 
stable solutions for arbitrary long times provided $\Delta t$ is chosen small enough. 
The boundary conditions
are imposed such that at the border for incoming momenta the Green  function is calculated
from the Eilenberger equation while for outgoing momenta the scattering condition Eq.(\ref{22}) is applied.
\begin{figure}
\includegraphics[width=0.5\textwidth]{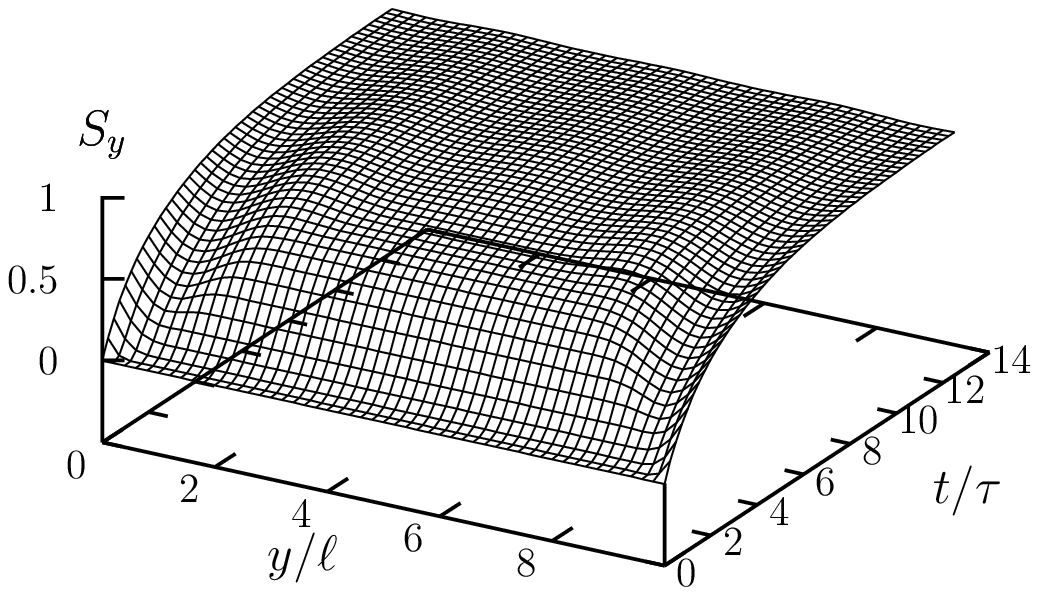}
\includegraphics[width=0.5\textwidth]{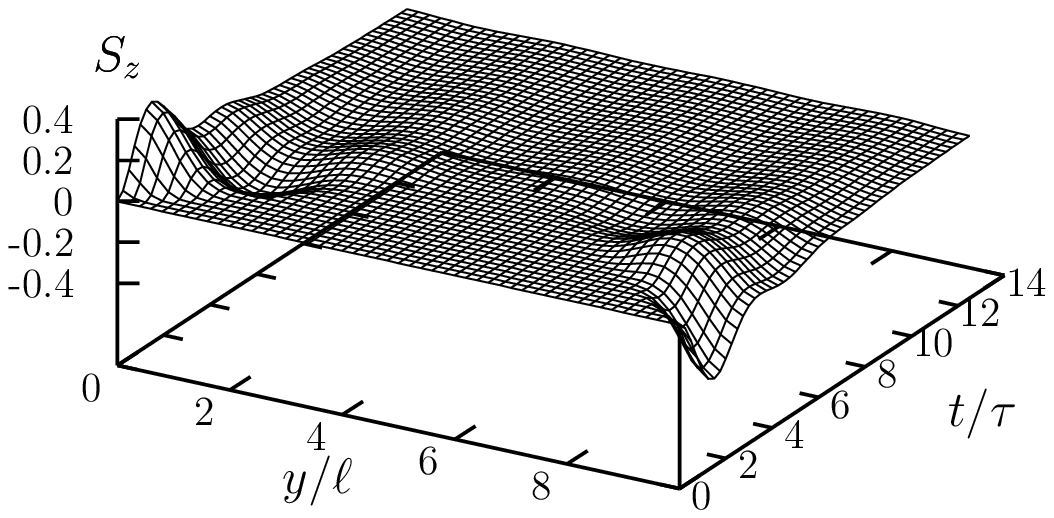} \newline
\includegraphics[width=0.5\textwidth]{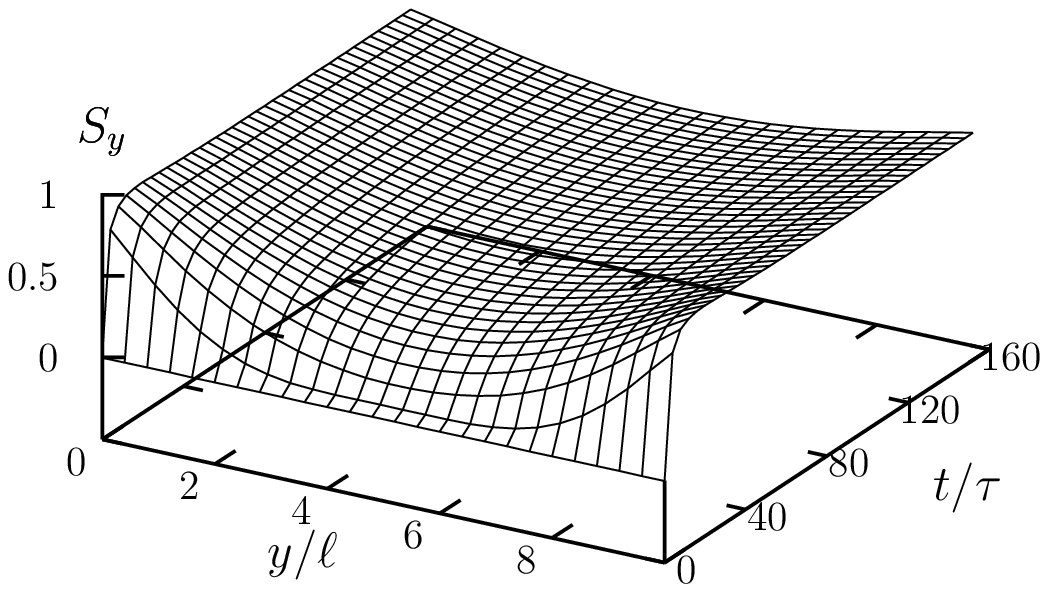}
\includegraphics[width=0.5\textwidth]{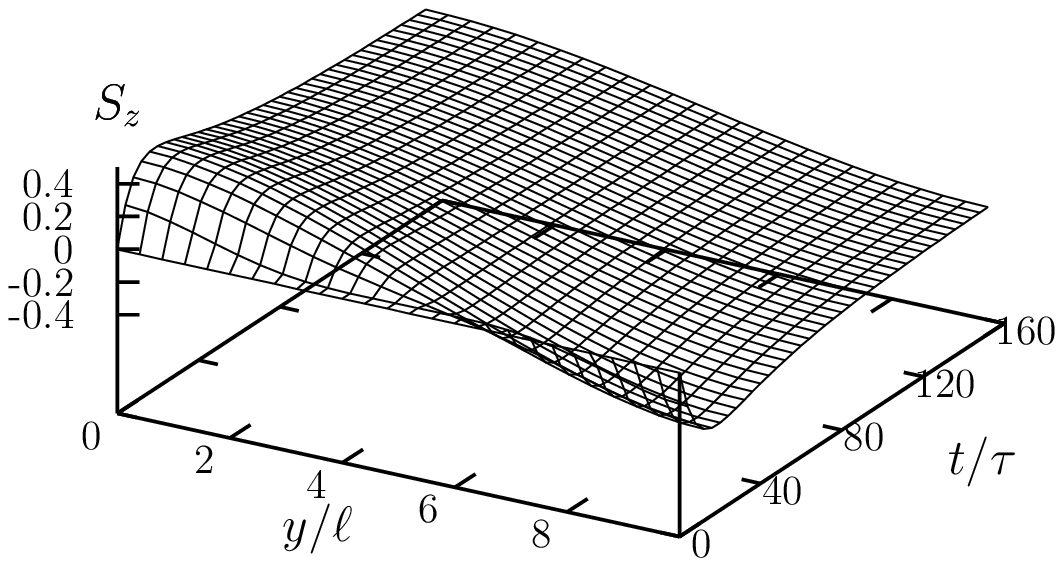} 
\caption{\label{fig2} Voltage induced spin polarization as function of $y$ and $t$
at $x = L_x/2$ on a strip of length $L_x = 20\,l$ and width $L_y = 10\,l$ for
$\alpha/v_F = 10^{-3}$. The upper figures are obtained for $\alpha p_F
\tau = 1$, whereas in the lower figures $\alpha p_F \tau =0.1$. Left panel: $s_y$, right panel: $s_z$,
both in units of the bulk polarization $s_0$.}
\end{figure}
Fig. \ref{fig2} shows the time evolution of the spin polarization on a line across the strip at
$x = L_x/2$, the Rashba parameter is $\alpha/v_F = 10^{-3}$. In the
upper part of the figure the elastic scattering rate $\tau$ is
chosen such that $\alpha p_F \tau = 1$ which is already beyond the reach of the diffusion equation approach.
The $y-$component $s_y$ of the spin polarization (left panel) increases monotonically from zero to the bulk
value $s_0$ while $s_z$ (right panel) shows strong oscillations close to the boundaries
which decay after a few periods. These oscillations are associated with a spin current
polarized in $z-$direction and flowing in $y-$direction. The magnitude of this spin current
corresponds to a spin Hall conductivity $\sigma_{sH} \simeq e/8\pi$ which however persists only on a short 
time scale after switching on the electrical field.  
In the stationary limit only the bulk polarization of $s_y$ survives while the polarization
of $s_x$ and $s_z$ is restricted to a small region around the corners
of the strip\cite{mishchenko2004,raimondi2006}.
In addition, the shape of these corner polarizations depends strongly on the details of the
coupling between the reservoirs and the strip.
In the lower part of Fig. \ref{fig2} we choose $\alpha p_F \tau =
0.1$, i.e. the spin dynamics is diffusive. 
Note that at the boundaries $s_y$ approaches $s_0$ on a much shorter
time scale than in the bulk, in accordance with the boundary condition
(\ref{31}). Due to the overdamped dynamics of the spins there are
neither in $s_y$ nor in $s_z$ time dependent oscillations.

To summarize, 
the quasiclassical approach provides a versatile theoretical framework for the
description of the coherent spin dynamics in confined electron
systems in the presence of spin-orbit coupling.

This work was supported by the Deutsche Forschungsgemeinschaft through
Sonderforschungsbereich 484.

\end{document}